# A short note on Reitlinger thermodynamic cycles

## Amelia Carolina Sparavigna

Department of Applied Science and Technology, Politecnico di Torino, Torino, Italy

**Abstract:** It is well known that Carnot cycle is the thermodynamic cycle which has the best thermal efficiency. However, an entire class of cycles exists that can have the same maximum efficiency: this class is that of the regenerative Reitlinger cycles. Here we discuss them.

**Keywords**: Thermodynamics, Thermodynamic cycles, regenerative cycles, Thermal efficiency.

Generally, the Carnot cycle is the only thermodynamic cycle that, during the lectures on physics, is discussed as having the maximum thermal efficiency. This approach yields the following result: it is less known that an entire class of cycles exists, having a cycle efficiency which is the same of the Carnot cycle. This is the class of the regenerative Reitlinger cycles. They consist of two isothermal and two polytropic processes of the same kind [1,2], so that the heat which is absorbed during a polytropic, is exactly the same that it is rejected on the other polytropic process. Therefore, if we have a perfect regeneration of heat, by means of which the heat rejected during the polytropic is transferred to a thermal storage (the regenerator) and then transferred back to the working fluid, the thermal efficiency of the Reitlinger cycle equals that of the Carnot cycle (in fact, it is a Reitlinger cycle too).
Of all the Reitlinger cycles, the Carnot cycle is unique in requiring the least regeneration, namely, none at all because its polytropics are adiabatics [1]. However, the mechanical work of the Carnot cycle is not the best we can obtain between extremal states, as we can see from the diagram in Figure 1, which is comparing Carnot and Stirling cycles, having the same temperature and volume extremes [1].

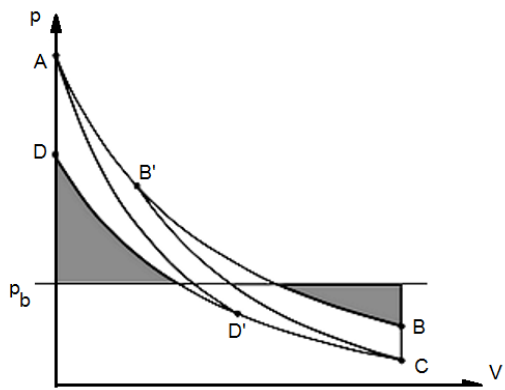

Figure 1: The figure (adapted from Ref.1) shows a Carnot cycle inscribed in a Stirling cycle in a p-V diagram. The optimum cosntant buffer pressure is also shown. The work of the Stirling cycle ABCD is greated than the work of Carnot cycle AB'CD'.

Let us note that the ideal Stirling cycle is also a Reitlinger cycle, having as polytropics two isochoric segments. It is the most popular example of a cycle having the same thermodynamic efficiency of the Carnot cycle; however, to attain this result, the Stirling cycle makes quite heavy demands on the process of regeneration [3].

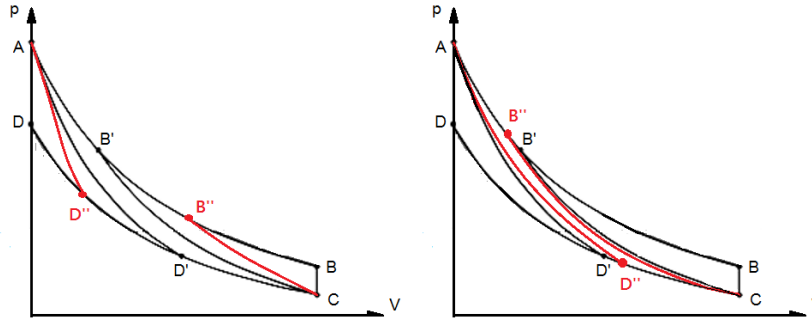

Figure 2: The figure shows how a Reitlinger cycle can be different from a Carnot cycle, in a p-V diagram. Working between the same isothermals, with the same thermal efficiency, a regenerative Reitlinger cycle AB''CD'' can give more or less work, depending on the polytropic process the cycle is performing between the same extremal states.

As observed in [6], there are ten elementary power cycles which follow from the combinations of five typical thermodynamic changes of state (Fig. 3). In the Figure we can see them and the names of their inventors (for other cycles, see [7]). In [6], Carnot, Ericsson and Stirling cycles are distinguished from the Reitlinger cycles, because they have a specific importance in thermodynamics. They are cycles containing the isothermal processes of compression and expansion, which have the most general form in the idealized cycles analyzed by Reitlinger [4,5]. As previously told, in these cycles we have, besides the two isothermal processes, two polytropic regenerative processes.

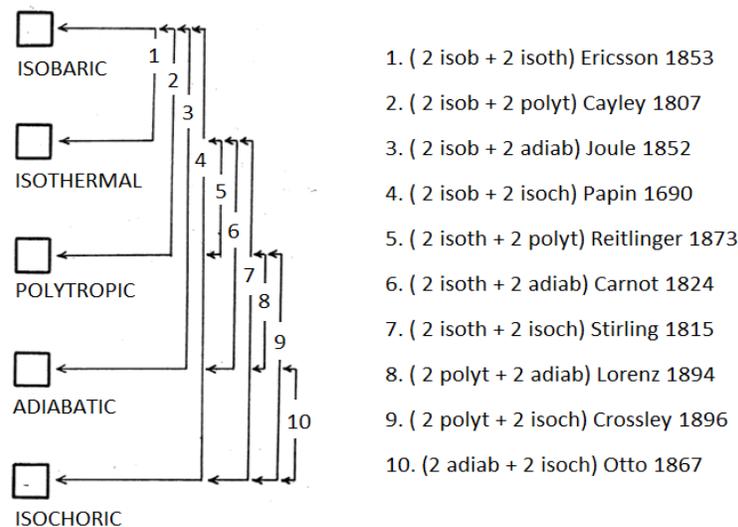

Figure 3: The elementary thermodynamic cycles (figure adapted from [6]).

For any thermodynamic cycle, reversible or irreversible, after one cycle, the working fluid is again in its initial state and thus the change of its internal energy is zero. In this manner, the first principle of thermodynamics tell us that the mechanical work produced by the cycle is the difference of input heat energy $Q_{in}$ minus the energy dissipated in waste heat $Q_{out}$. Heat engines transform thermal energy into mechanical energy or work, $W$, so that $W = Q_{in} - Q_{out}$. We can calculate the thermal efficiency of the cycle as the dimensionless performance measure of the use of thermal energy. The thermal efficiency of a heat engine is the percentage of heat energy that is transformed into work, so that:

(1) $$\eta = \frac{W}{Q_{in}}$$

For a Carnot engine, it is $\eta = 1 - T_C/T_H$, where $T_H, T_C$ are the temperatures of the furnace and of the cold sink. Let us discuss the thermal efficiency of the Stirling cycle. Using a p-V diagram, the cycle appears as in the Figure 4. In the same figure, the Ericsson cycle and Reitlinger cycle are also shown.

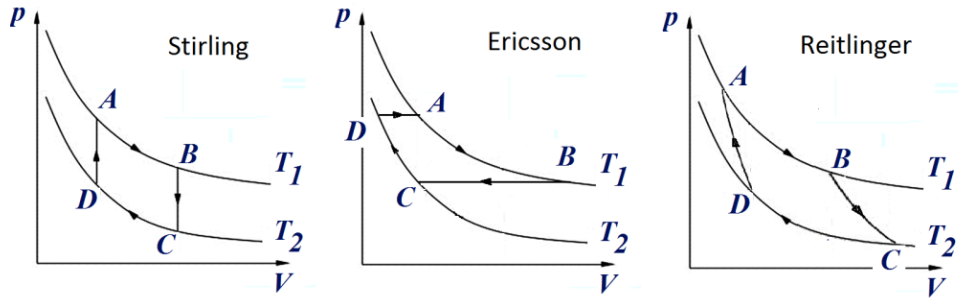

Figure 4: Stirling, Ericsson and Reitlinger cycles in p-V diagrams.

The work can be easily calculated as:

(2) $$W = nR(T_1 - T_2)\ln\frac{V_B}{V_A}$$

In (2), $n$ is the number of moles and $R$ the universal gas constant. Heat is gained by the thermodynamic system from the reversible isochoric transformation from D to A and during the isothermal path AB. During isochoric process, the heat gained is: $Q_{isoc} = nC_V(T_1 - T_2)$. $C_V$ is the molar specific heat for an isochoric process. During isothermal process, the heat gained is $Q_{isot} = nRT_1 \ln(V_B/V_A)$.

Let us note that, during the isochoric process, the fluid is obtaining heat from an infinite number of thermal reservoirs [8]. This same amount of heat is lost during the isochoric cooling process, with a thermal exchange with the same reservoirs. Then, for each of the infinite thermal reservoirs that we meet during the isochoric process, it happens what we see in the Figure 5. In this figure, we have two thermal machines that must have the same efficiency, to satisfy the second principle of thermodynamics.

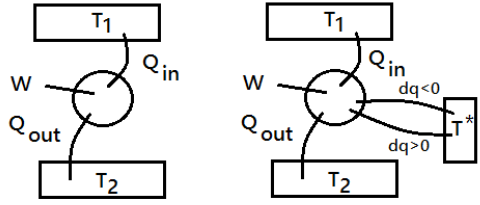

Figure 5: The two reversible cycles in the figure have the same efficiency. If it were not so, we should violate the second principle of thermodynamics. Let us suppose the efficiency of the right machine larger than that of the left one. Let us consider the same work *W* produced by the two machines, and operate the machine on the left in reversed manner. It is easy to see that the net result of these two machines operating together is that of transferring some heat from the low temperature reservoir to the high temperature reservoir, violating the Clausius statement of the second principle of thermodynamics.

From an engineering perspective, the regenerator simply stores the heat and therefore the abovementioned thermal reservoirs are not involved. Consequently, considering the system made of working fluid and regenerator, the thermal efficiency is:

$$(3) \quad \eta = \frac{W}{Q} = \frac{nR(T_1 - T_2)\ln\frac{V_B}{V_A}}{nRT_1 \ln\frac{V_B}{V_A}} = 1 - \frac{T_2}{T_1}$$

In (3), *Q* is the heat the system receives during the high temperature isothermal process. This efficiency is equal to that of a Carnot cycle which is working between the same two isotermal processes. We can repeat the calculation for the Ericsson cycle. The work is:

$$(4) \quad \begin{aligned} W &= nC_P(T_1 - T_2) + nRT_1 \ln\frac{V_B}{V_A} - nC_P(T_1 - T_2) + nRT_2 \ln\frac{V_D}{V_C} \\ &= nRT_1 \ln\frac{V_B}{V_A} - nRT_2 \ln\frac{nRT_2}{p_C}\frac{p_D}{nRT_2} = nRT_1 \ln\frac{V_B}{V_A} - nRT_2 \ln\frac{p_A}{p_B} \\ &= nRT_1 \ln\frac{V_B}{V_A} - nRT_2 \ln\frac{V_B}{V_A} \end{aligned}$$

In (4), *Cp* is the molar specific heat at constant pressure. It is clear that the heat lost and gained during the two isobaric processes is the same. Therefore, the thermal efficiency, in the case of a perfect regeneration, is given by:

$$(5) \quad \eta = \frac{W}{Q} = \frac{nR(T_1 - T_2)\ln\frac{V_B}{V_A}}{nRT_1 \ln\frac{V_B}{V_A}} = 1 - \frac{T_2}{T_1}$$

In (5), $Q$ is the heat the system receives during the high temperature isothermal process. Let us conclude with a Reitling cycle, where polytropics are given by equations $pV^\alpha = const$ and $TV^{\alpha-1} = const$. The molar specific heat of such polytropic process is $C_\alpha$. Let us note that from polytropic equation we have (see Figure 4):

(6)
$$T_1 V_A^{\alpha-1} = T_2 V_D^{\alpha-1}$$
$$T_1 V_B^{\alpha-1} = T_2 V_C^{\alpha-1}$$

Therefore, we have:

(7)
$$\frac{V_A^{\alpha-1}}{V_B^{\alpha-1}} = \frac{V_D^{\alpha-1}}{V_C^{\alpha-1}} \rightarrow \frac{V_A}{V_B} = \frac{V_D}{V_C}$$

Then:

(8)
$$W = nC_\alpha(T_1 - T_2) + nRT_1 \ln\frac{V_B}{V_A} - nC_\alpha(T_1 - T_2) + nRT_2 \ln\frac{V_D}{V_C}$$
$$= nRT_1 \ln\frac{V_B}{V_A} - nRT_2 \ln\frac{V_B}{V_A}$$

Again, we find a thermal efficiency of the system (fluid and regenerator), which is again equal to that of the Carnot engine. Therefore, since the polytropic index α can have any value, we have an infinite number of thermodynamic cycles that have the same value of thermal efficiency, equal to that of the Carnot cycle when operating between the same two isothermal processes. Let us stress that these cycles incorporate a regenerative heat transfer process, in place of adiabatic compression and expansion of the Carnot cycle [5], or, if preferred, an infinite number of monothermal processes, not influencing the efficiency of the cycle. Moreover, it is better to also remark that the fact of possessing the same thermal efficiency does not mean that the same work is obtained from different reversible cycles, when they are operating between the same extremal states.